\documentclass[
 reprint,
 superscriptaddress,
 amsmath,
 amssymb,
 aps,
 prl,
]{revtex4-2}

\usepackage{graphicx}
\usepackage{dcolumn}
\usepackage{bm}
\usepackage{xcolor}   
\usepackage{kotex}
\usepackage{tabularx} 
\usepackage[normalem]{ulem}
\usepackage{overpic}
\usepackage{hyperref}
\hypersetup{
    colorlinks=true,
    linkcolor=blue,
    filecolor=magenta,      
    urlcolor=cyan,
    citecolor=blue,
}

\begin{document}
\preprint{APS/123-QED}

\title{Search for Axion Dark Matter around \texorpdfstring{$15.6\,\mu\text{eV}$}{15.6 microeV} with a Broadly Tunable High-Temperature Superconducting Cavity}

\author{Jiwon Lee}
\affiliation{Department of Physics, Korea Advanced Institute of Science and Technology (KAIST), Daejeon 34141, Republic of Korea}
\affiliation{Dark Matter Axion Group, Institute for Basic Science (IBS-DMAG), Daejeon 34126, Republic of Korea}

\author{Jinsu Kim}
\affiliation{Dark Matter Axion Group, Institute for Basic Science (IBS-DMAG), Daejeon 34126, Republic of Korea}

\author{Ohjoon Kwon}
\email[Corresponding author: ]{o1tough@ibs.re.kr}
\affiliation{Dark Matter Axion Group, Institute for Basic Science (IBS-DMAG), Daejeon 34126, Republic of Korea}

\author{Saebyeok Ahn}
\affiliation{Dark Matter Axion Group, Institute for Basic Science (IBS-DMAG), Daejeon 34126, Republic of Korea}

\author{Seonjeong Oh}
\affiliation{Dark Matter Axion Group, Institute for Basic Science (IBS-DMAG), Daejeon 34126, Republic of Korea}

\author{Sergey Uchaikin}
\affiliation{Dark Matter Axion Group, Institute for Basic Science (IBS-DMAG), Daejeon 34126, Republic of Korea}

\author{Boris Ivanov}
\affiliation{Dark Matter Axion Group, Institute for Basic Science (IBS-DMAG), Daejeon 34126, Republic of Korea}

\author{Arjan Ferdinand van Loo}
\affiliation{Dark Matter Axion Group, Institute for Basic Science (IBS-DMAG), Daejeon 34126, Republic of Korea}

\author{Yasunobu Nakamura}
\affiliation{Department of Applied Physics, Graduate School of Engineering,\\
The University of Tokyo, Bunkyo-ku, Tokyo 113-8656, Japan}
\affiliation{RIKEN Center for Quantum Computing (RQC), Wako, Saitama 351-0198, Japan}

\author{Dojun Youm}
\affiliation{Dark Matter Axion Group, Institute for Basic Science (IBS-DMAG), Daejeon 34126, Republic of Korea}

\author{SungWoo Youn}
\affiliation{Dark Matter Axion Group, Institute for Basic Science (IBS-DMAG), Daejeon 34126, Republic of Korea}

\author{HyoungSoon Choi}
\affiliation{Department of Physics, Korea Advanced Institute of Science and Technology (KAIST), Daejeon 34141, Republic of Korea}

\date{\today}

\begin{abstract}
We report an axion dark matter search in the mass range of $15.30\text{--}15.85\,\mu\text{eV}$ using a broadly tunable high-temperature superconducting (HTS) haloscope. By soldering substrate-stripped rare-earth barium copper oxide (REBCO) films onto a copper shell, we achieved a quality factor 3--4 times higher than in copper cavities across its tuning band in an 8\text{-T} magnetic field. We set robust Frequentist 90\%~confidence-level exclusion limits on the axion--photon coupling down to 1.3 times the KSVZ coupling, while a complementary Bayesian analysis achieves KSVZ-level sensitivity.
\end{abstract}

\maketitle

The experimental pursuit of the quantum chromodynamics (QCD) axion---a well-motivated dark matter candidate emerging from the Peccei-Quinn solution to the strong $CP$ problem~\cite{Peccei1977, Weinberg1978, Wilczek1978}---requires navigating an immense parameter space of unknown mass and coupling~\cite{Planck2020, Rubin1970, PDG2024, Abel2020nEDM}. Cavity haloscopes address this by exploiting axion-to-photon conversion inside a microwave resonator permeated by a strong magnetic field~\cite{sikivie1983experimental}. The experimental search rate $d\nu/dt$---the figure of merit for haloscopes---scales as $B^4 V^2 C^2 Q / T_{\text{sys}}^2$~\cite{Dicke1946, Kim2020JCAP}, measuring search performance across magnetic field $B$, cavity volume $V$, form factor $C$, quality factor $Q$, and system noise $T_{\text{sys}}$. Because modern experiments routinely employ high-field superconducting magnets and quantum-noise-limited readout chains to optimize $B$ and $T_{\text{sys}}$~\cite{Kwon2021FirstResults, Kim2022UltraLowJPA}, discovery speeds are now fundamentally restricted by normal-metal cavity dissipation, limiting cryogenic copper resonators near $Q \lesssim 10^5$~\cite{anormalous1950nat}. 

Recently, high-temperature superconducting (HTS) coated conductors based on biaxially-textured rare-earth barium copper oxide (REBCO) have emerged as a compelling solution. Their exceptionally high upper critical fields ($H_{c2} > 100\,\text{T}$) prevent field-induced suppression of superconductivity, while dense flux-pinning defects suppress high-frequency vortex dissipation \cite{Wu1987Hc2, Obradors2014Review, Pompeo2008Vortex, Romanov2020REBCORs}. Pioneering HTS cavities demonstrated that aligning the longitudinal surface currents~$J_z$ of the symmetric $\text{TM}_{010}$ mode in a cylindrical cavity with inter-tape assembly joints forms parallel-plate waveguides below cutoff, yielding fixed-frequency quality factors up to $Q_0 \sim 1.4 \times 10^7$ at $8\,\text{T}$~\cite{Ahn2022PRApplied, Ahn2026arxiv}. However, converting this high-$Q$ capability into a practical, tunable haloscope presents a significant electrodynamic challenge. Introducing a conventional off-axis tuning rod breaks the azimuthal symmetry of the $\text{TM}_{010}$ mode. This symmetry breaking induces transverse electric-field components $E_\phi$ across the longitudinal seams, transforming the below-cutoff gaps into unattenuated radiation channels that severely degrade the quality factor.

We resolve this tuning leakage problem by establishing full electrical continuity across substrate-stripped REBCO films. Extending thin-sheet ultra-light cavity (ULC) engineering~\cite{Yi2023DFSZ, ahn2024prx}, we performed face-to-face indium soldering of copper-backed REBCO films onto the inner wall of a copper shell. This continuous conductive backing short-circuits the transverse $E_\phi$ fields, eliminating transverse electromagnetic (TEM) radiation leakage during off-axis tuning-rod rotation.

In this Letter, we report an axion haloscope search, DMAG-8T---building upon our previous 8-T haloscope searches~\cite{Kwon2021FirstResults, Kim2022UltraLowJPA}---in the mass range of $15.30\text{--}15.85\,\mu\text{eV}$ ($3.700\text{--}3.833\,\text{GHz}$). By utilizing an HTS cavity continuously tunable across $3.21\text{--}4.05\,\text{GHz}$ that maintains an unloaded quality factor 3--4 times higher than in copper cavities in an $8\,\text{T}$ magnetic field, and integrating it with a near-quantum-noise-limited Josephson parametric amplifier (JPA)~\cite{Yamamoto2008JPA,Kutlu2021JPA}, we significantly accelerated the search rate, surpassing previous direct search limits in this mass range by a factor of 60~\cite{RBF1987, *RBF1989}. We set Frequentist exclusion limits on the axion--photon coupling $g_{a\gamma\gamma}$ down to 1.3 times the KSVZ coupling~\cite{Kim1979, Shifman1980}, while a complementary Bayesian analysis achieves KSVZ-level sensitivity ($g_{a\gamma\gamma} \approx g_{\text{KSVZ}}$)~\cite{Palken2020BPM}.


\begin{figure}[t]
\centering
\includegraphics[width=\columnwidth]{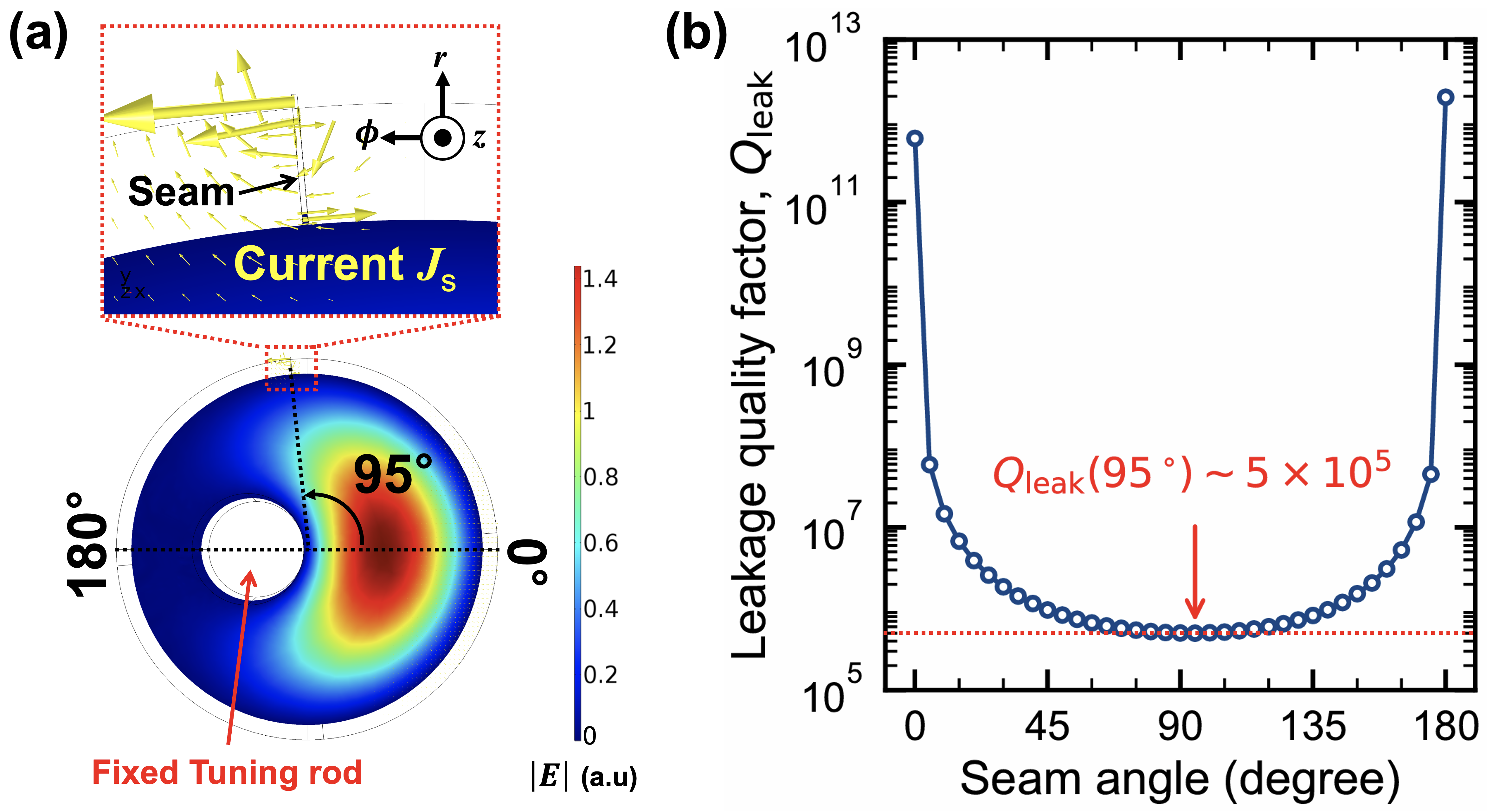}
\caption{Simulated TEM radiation leakage induced by a symmetry-breaking tuning rod. (a)~2D electric field distribution of the $\text{TM}_{010}$ mode in an asymmetric cavity. The zoomed inset highlights transverse surface current~$\bm{J}_s$ crossing the seam gap, which excite cutoff-free TEM modes and drive RF power leakage outward along the radial direction~$r$. (b)~Simulated single-gap radiation quality factor $Q_{\text{leak}}$ assuming a perfect electric conductor boundary as a function of a single gap's azimuthal angle, measured relative to the axis connecting the cavity and the fixed off-axis tuning rod centers. Except for collinear symmetric orientations ($0^{\circ}$ and $180^{\circ}$), field asymmetry unavoidably forces transverse surface currents across the seam, causing $Q_{\text{leak}}$ to drop drastically across almost all angles down to a minimum of $5 \times 10^5$ at $95^{\circ}$. In a realistic multi-seam cavity, the cumulative dissipation across all open seams severely caps the operational quality factor to normal-metal levels.}
\label{fig:gap_sim}
\end{figure}
High-performance HTS cavities are traditionally constructed using longitudinal wedges clad with biaxially textured REBCO tapes to match the longitudinal surface current paths $J_z$ of the $\text{TM}_{010}$ mode~\cite{Ahn2022PRApplied, Ahn2026arxiv}. For this longitudinal polarization $E_z$, assembly gaps of width $w \lesssim 20\,\mu\text{m}$ act as parallel-plate waveguides operating in the transverse electric $\text{TE}_{10}$ mode far below the cutoff frequency ($\nu_c \approx c/2w \sim 7.5\text{--}15\,\text{THz} \gg 3.7\,\text{GHz}$). The evanescent field decays exponentially along the radial gap depth $r$ as $\exp(-\pi r / w)$~\cite{Jackson1999,Pozar2011}. Mechanically delaminating and stripping the lossy $50\text{-}\mu\text{m}$-thick Hastelloy substrate eliminates the primary Ohmic dissipation channel behind the gap. By utilizing the $20\text{-}\mu\text{m}$-thick low-loss copper stabilizer as a conductive backing, the evanescent field decays completely within the copper-backed waveguide depth, effectively suppressing seam dissipation. This architecture has yielded fixed-frequency quality factors up to $Q_0 \sim 1.3 \times 10^7$ at $8\,\text{T}$~\cite{Ahn2026arxiv} in non-tunable resonators.

However, converting this fixed-frequency capability into a practical, wideband-tunable haloscope presents a severe electrodynamic challenge. Introducing a conventional off-axis tuning rod
fundamentally breaks the azimuthal field symmetry~\cite{RBF1987, Kwon2021FirstResults, Kim2022UltraLowJPA, Yi2023DFSZ, *ahn2024prx, ADMX2004, *ADMX2018, *ADMX2020, *ADMX2021, *ADMX2025a, *ADMX2025b}. This asymmetry excites transverse electric fields $E_\phi$ across the assembly gaps. Because parallel-plate gaps impose no cutoff frequency for transverse electromagnetic (TEM) modes, extending the radial gap depth becomes insufficient; the $E_\phi$ fields drive unattenuated TEM radiation leakage into the lossy substrates. Finite-element simulations using COMSOL~\cite{COMSOL} confirm that field asymmetry forces transverse surface currents $J_\phi$ across the seam [Fig.~\ref{fig:gap_sim}(a)]. Except when a gap lies on the mirror-symmetry plane connecting the cavity and rod centers ($0^\circ$ and $180^\circ$), this cutoff-free leakage drastically reduces the single-gap quality factor $Q_{\text{leak}}$. Due to rod-induced field distortion, $Q_{\text{leak}}$ reaches its minimum of $5\times 10^5$ near perpendicular alignment [Fig.~\ref{fig:gap_sim}(b)]. For realistic cavities with $N \sim 12\text{--}32$ seams~\cite{Ahn2022PRApplied,Ahn2026arxiv}, the total quality factor combines cumulatively:
\begin{equation}
\frac{1}{Q_{\text{total}}} = \frac{1}{Q_0} + \sum_{i=1}^N \frac{1}{Q_{\text{leak},i}},
\label{eq:Qtotal}
\end{equation}
collapsing the operational quality factor down to conventional copper-cavity levels.

\begin{figure}[t]
\centering
\includegraphics[width=\columnwidth]{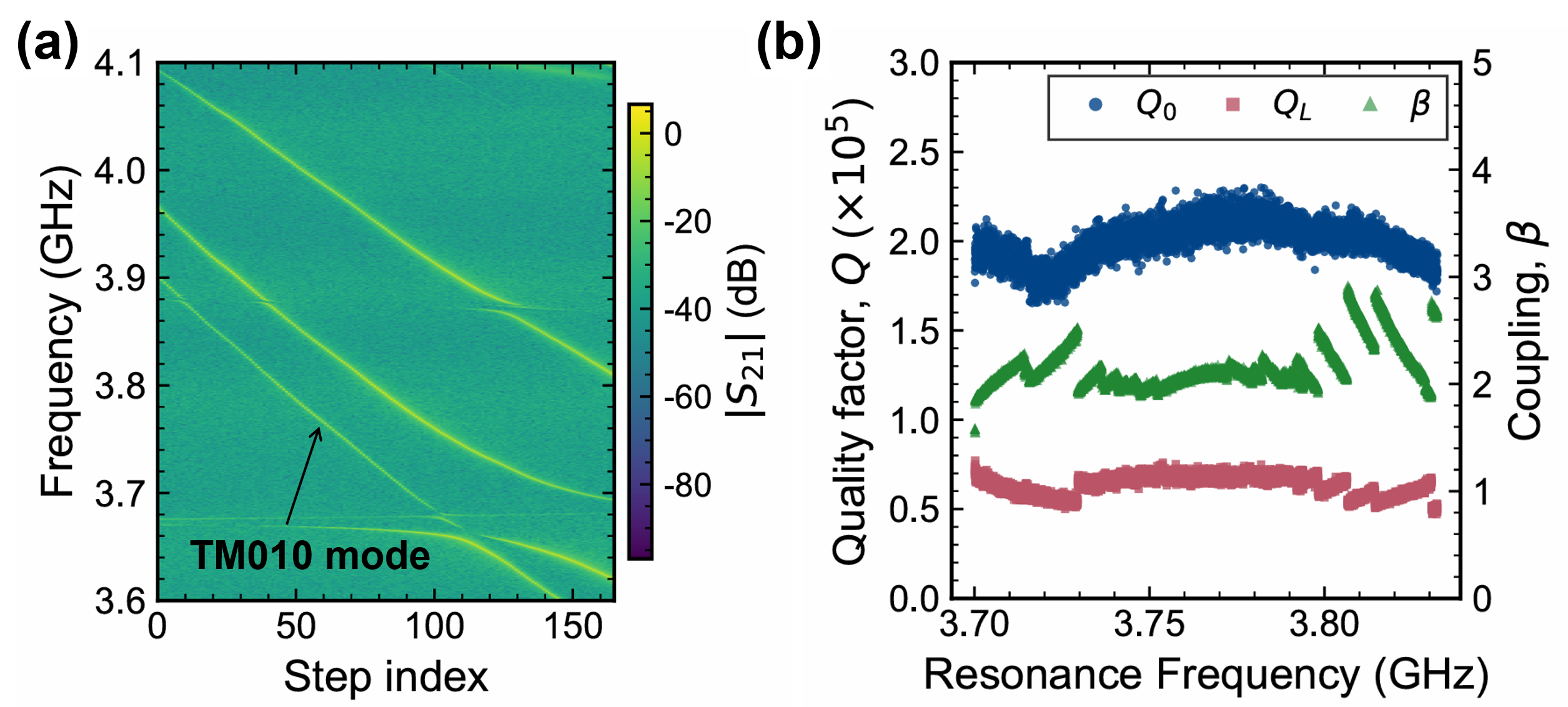}
\caption{Measured characteristics of the HTS cavity for DMAG-8T under an $8\,\text{T}$ magnetic field. (a)~Measured mode-map across the $3.6\text{--}3.9\,\text{GHz}$ tuning band, showing smooth tracking of the $\text{TM}_{010}$ mode (black arrow). (b)~Unloaded quality factor~$Q_0$, loaded quality factor~$Q_L$, and coupling strength~$\beta$ as a function of resonance frequency, measured in situ during the physics run. The discrete jumps in $Q_L$ and $\beta$ reflect deliberate adjustments of the antenna insertion depth via a linear piezo to maintain optimal coupling efficiency.}
\label{fig:cavity_perf}
\end{figure}

We eliminate this TEM leakage channel by ULC engineering~\cite{Yi2023DFSZ,ahn2024prx}. To maximize volume utilization within the DMAG-8T magnet bore for axion searches around $3.7\,\text{GHz}$, the physical cavity was constructed with an inner diameter of $103\,\text{mm}$ and a height of $220\,\text{mm}$ ($V_{\text{eff}} \approx 1.68\,\text{L}$). All inner rf-exposed surfaces---including the cavity sidewalls, endcaps, and the tuning rod's outer boundary---were clad with substrate-stripped EuBCO tapes, ensuring that the electromagnetic field sees a continuous superconducting boundary. Substrate removal was accomplished via strain-controlled mechanical delamination along the weak dielectric buffer layer, stripping lossy $50\text{-}\mu\text{m}$-thick Hastelloy substrates while leaving the $2.5\text{-}\mu\text{m}$ EuBCO film supported by its $1.5\text{-}\mu\text{m}$ silver protection and $20\text{-}\mu\text{m}$-thick copper stabilizer layers~\cite{vanDerLaan2010Strain, Shin2014Transverse, Ahn2026arxiv}. 

To assemble this 3D cavity without degrading the superconducting film, we implemented a two-stage thermal soldering process. In the first stage, a $0.5\text{-mm}$-thick oxygen-free high-conductivity (OFHC) copper sheet was pre-rolled into a cylindrical pipe. Substrate-stripped and copper-backed EuBCO tapes were precisely aligned along the inner surface of the copper pipe, and full-surface heating at $T_{\text{melt}} \approx 156\,^{\circ}\text{C}$ achieved face-to-face indium soldering, metallurgically bonding the exposed silver/copper film backings directly to the copper shell. Tapes were similarly soldered onto the OFHC copper endcaps, as well as the tuning rod assembly (a $30\text{-mm}$-diameter copper pipe and its caps) in a single indium-soldering step. In the second stage, with the tuning rod installed inside, the HTS-clad cavity sidewall and endcaps were joined along their assembly seams using a low-melting-point indium-bismuth (In-Bi) alloy (66.3\% In and 33.7\% Bi; $T_{\text{melt}} \approx 72\,^{\circ}\text{C}$). This lower-temperature step securely seals the cavity body without reflowing or thermally damaging the primary indium-bonded HTS layers.

Frequency tuning is actuated by mounting the HTS-clad tuning rod on a rotation axis offset $8\,\text{mm}$ from the cavity center ($16\text{-}\text{mm}$ eccentric orbit), driven by an Attocube piezo rotator~\cite{Attocube} with a 1:5 brass-gear reduction module covering $3.21\text{--}4.05\,\text{GHz}$. A bottom copper pivot pin in a retaining well provides direct thermal contact, thermalizing the rod and suppressing thermal noise.

To model residual dissipation in the absence of TEM leakage ($\sum 1/Q_{\text{leak},i} \to 0$), we evaluate the unloaded quality factor ($Q_0$), which is governed by cavity-wall surface losses ($Q_0 \approx Q_{\text{surface}}$):
\begin{equation}
\frac{1}{Q_0} \approx \frac{1}{Q_{\text{surface}}} = \frac{1}{Q_{\text{HTS}}} + \frac{1}{Q_{\text{seam}}} + \frac{1}{Q_{\text{normal}}},
\label{eq:Q0_loss}
\end{equation}
where $1/Q_{\text{HTS}}$ represents intrinsic surface dissipation in the strongly pinned EuBCO film (which is exceptionally low at $8\,\text{T}$), and $1/Q_{\text{seam}}$ accounts for ohmic losses from exposed indium and In-Bi joint alloys along assembly seams. Dielectric volume loss from the sapphire tuning axle ($Q_{\text{dielectric}} \sim 10^9$) is orders of magnitude smaller than $Q_0$ and strictly negligible. The normal-metal term~$1/Q_{\text{normal}}$ incorporates dissipation from the copper pivot pin, antenna coupling hole, and localized HTS degradation. Post-run inspection upon cavity disassembly revealed that incomplete cleaning of soldering flux residues caused localized oxidation of the REBCO lattice, partially exposing underlying silver protection and copper stabilizer layers. Remarkably, despite these localized surface defects, the complete suppression of TEM leakage demonstrates the structural robustness of our seamless ULC architecture, continuously tracking the $\text{TM}_{010}$ mode [Fig.~\ref{fig:cavity_perf}(a)] while enabling the cavity to reliably maintain $Q_0 \approx 200,000$ across $3.7\text{--}4.0\text{-GHz}$ at $8\,\text{T}$ [Fig.~\ref{fig:cavity_perf}(b)].

\begin{figure}[t]
\centering
\includegraphics[width=\columnwidth]{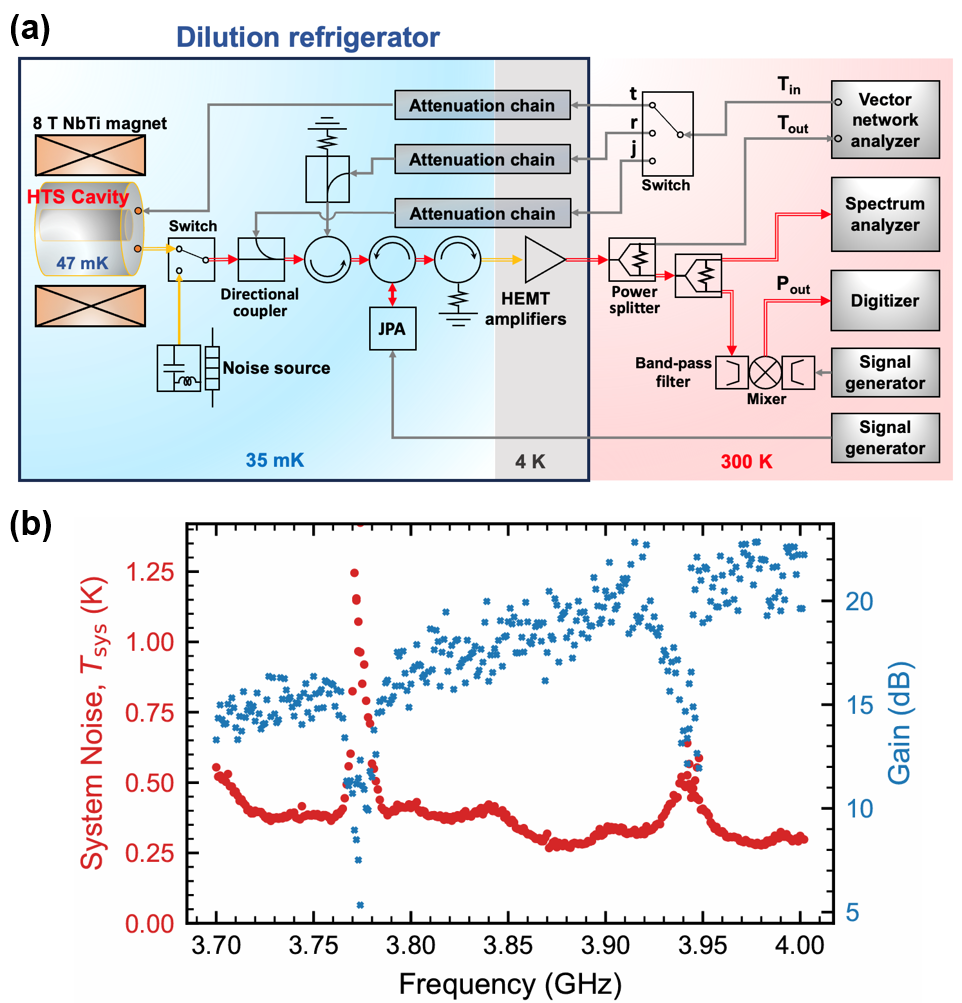}
\caption{RF receiver chain and active noise performance. (a)~Schematic of the DMAG-8T microwave readout chain (based on our previous setup~\cite{Kwon2021FirstResults, Kim2022UltraLowJPA}). Basic components (directional couplers, circulators, band-pass filters, and HEMT amplifiers) were replaced for $2\text{--}4\,\text{GHz}$ operation, while a flux-driven JPA at $35\text{-}\text{mK}$ stage targeted for the $3.7\text{--}4.0\text{-GHz}$ region was installed. Yellow arrows indicate NbTi superconducting coaxial lines, gray/red arrows represent Cu/CuNi coaxial lines and room-temperature rf paths, and switching ports $t$, $r$, $j$ select transmission, reflection, and JPA flux-bias characterization modes, respectively. 
(b)~Total system noise temperature $T_{\text{sys}}$ (red, left axis) and corresponding phase-insensitive JPA gain $G_{\text{JPA}}$ (blue, right axis) across $3.7\text{--}4.0\,\text{GHz}$. The system achieves a minimum $T_{\text{sys}} \approx 280\,\text{mK}$~($\sim 1.5\,\text{quanta}$) at $3.88\,\text{GHz}$.}
\label{fig:jpa_on}
\end{figure}

For DMAG-8T operation, the broad-band HTS cavity was thermalized to the mixing chamber plate ($T \approx 35\,\text{mK}$) of a Bluefors dilution refrigerator~\cite{Bluefors} operated within our established 8-T haloscope facility~\cite{Kwon2021FirstResults, Kim2022UltraLowJPA}, as schematically illustrated in Fig.~\ref{fig:jpa_on}(a). To accommodate the $3.7\text{--}4.0\text{-}\text{GHz}$ search band, all basic passive and active microwave components---including directional couplers, circulators, band-pass filters, and the cryogenic high electron mobility transistor (HEMT) amplifiers---were replaced for $2\text{--}4\text{-GHz}$ S-band operation while preserving the low-loss coaxial switching topology (ports $t, r, j$). The baseline added noise of the downstream cryogenic receiver line ($T_{\text{rx}} \approx 2.5\text{--}3.0\,\text{K}$) was calibrated via the standard $Y$-factor method with the JPA deactivated using a cryogenic thermal noise source.

Activating the flux-driven JPA---fabricated at RIKEN~\cite{Yamamoto2008JPA,Kim2022UltraLowJPA} and deployed specifically to provide optimal noise performance in the $3.7\text{--}4.0$-GHz target region---in phase-insensitive mode ($G_{\text{JPA}} \approx 15\text{--}18\,\text{dB}$) reduces the total system noise temperature $T_{\text{sys}}$, evaluated via the spectrum comparison method as:
\begin{equation}
T_{\text{sys}} = \frac{(T_{\text{rx}} + T_{\text{MXC}}^{\text{eff}}) \, r}{\eta_c G_{\text{JPA}}},
\label{eq:Tsys}
\end{equation}
where $r$ is the JPA-on to JPA-off power ratio, $T_{\text{MXC}}^{\text{eff}}$ is the effective thermal noise of the mixing chamber, and $\eta_c$ is the cavity coupling efficiency. As shown in Fig.~\ref{fig:jpa_on}(b), $T_{\text{sys}}$ reaches a minimum of $280\,\text{mK}$ ($\sim 1.5\,\text{quanta}$) at $3.88\,\text{GHz}$ and remains well below $400\,\text{mK}$ ($\lesssim 2.1\,\text{quanta}$) across most of the search band. Localized $T_{\text{sys}}$ spikes observed around $3.77$ and $3.94\,\text{GHz}$ stem from mode hybridization between the primary JPA resonance and stray parasitic modes localized on the JPA's internal superconducting island, suppressing parametric gain and leaving downstream HEMT noise unsuppressed.

To maintain optimal operation during continuous tuning, we implemented an automated JPA tracking routine. Following vector network analyzer (VNA) cavity characterization---during which the JPA is detuned to prevent saturation---flux sweeps identify the new resonance. The peak current $I_0$ is extracted using a Savitzky-Golay (SG) filter~\cite{Savitzky1964} and a Lorentzian fit:
\begin{equation}
G(I) = \frac{A}{1 + \left(\frac{I - I_0}{\gamma/2}\right)^2} + C.
\label{eq:lorentzian_gain}
\end{equation}
Pump power is subsequently adjusted to lock the gain at $15\text{--}18\,\text{dB}$, minimizing $T_{\text{sys}}$ and reducing dead time between tuning steps.

Continuous data acquisition also required mitigating a magneto-mechanical stall of the HTS tuning rod. Pinned flux in the EuBCO tapes induces a macroscopic magnetic moment $\bm{M}$, which generates a restoring torque $\bm{\tau} = \bm{M} \times \bm{B}_{\text{ext}}$ that inhibits rotation at $8\,\text{T}$. We bypassed this limitation via an automated field-cycling protocol: transiently ramping the field down to $7.7\,\text{T}$ (or $7.4\,\text{T}$) accelerates flux creep via the induced $d|\bm{B}_\mathrm{ext}|/dt$, rapidly decaying $\bm{M}$ and relieving the torque. Once the motor completes the rotation, the field is restored to $8.0\,\text{T}$ and a pre-flight check ($T_{\text{cav}} < 47\,\text{mK}$) verifies thermal stability, seamlessly transitioning the system into continuous physics integration.

The DMAG-8T search operated from December 25, 2025, to April 8, 2026, scanning $133\,\text{MHz}$ across $3.700\text{--}3.833\,\text{GHz}$ ($15.30\text{--}15.85\,\mu\text{eV}$). The data acquisition sequence stepped the frequency in $20\text{-kHz}$ increments ($\sim \Delta f_{\text{cav}}/3$), integrating power spectra for $780\,\text{s}$ per step and achieving a scan rate of $1.28\,\text{MHz/day}$.

\begin{figure*}[t]
\centering
\includegraphics[width=1.8\columnwidth]{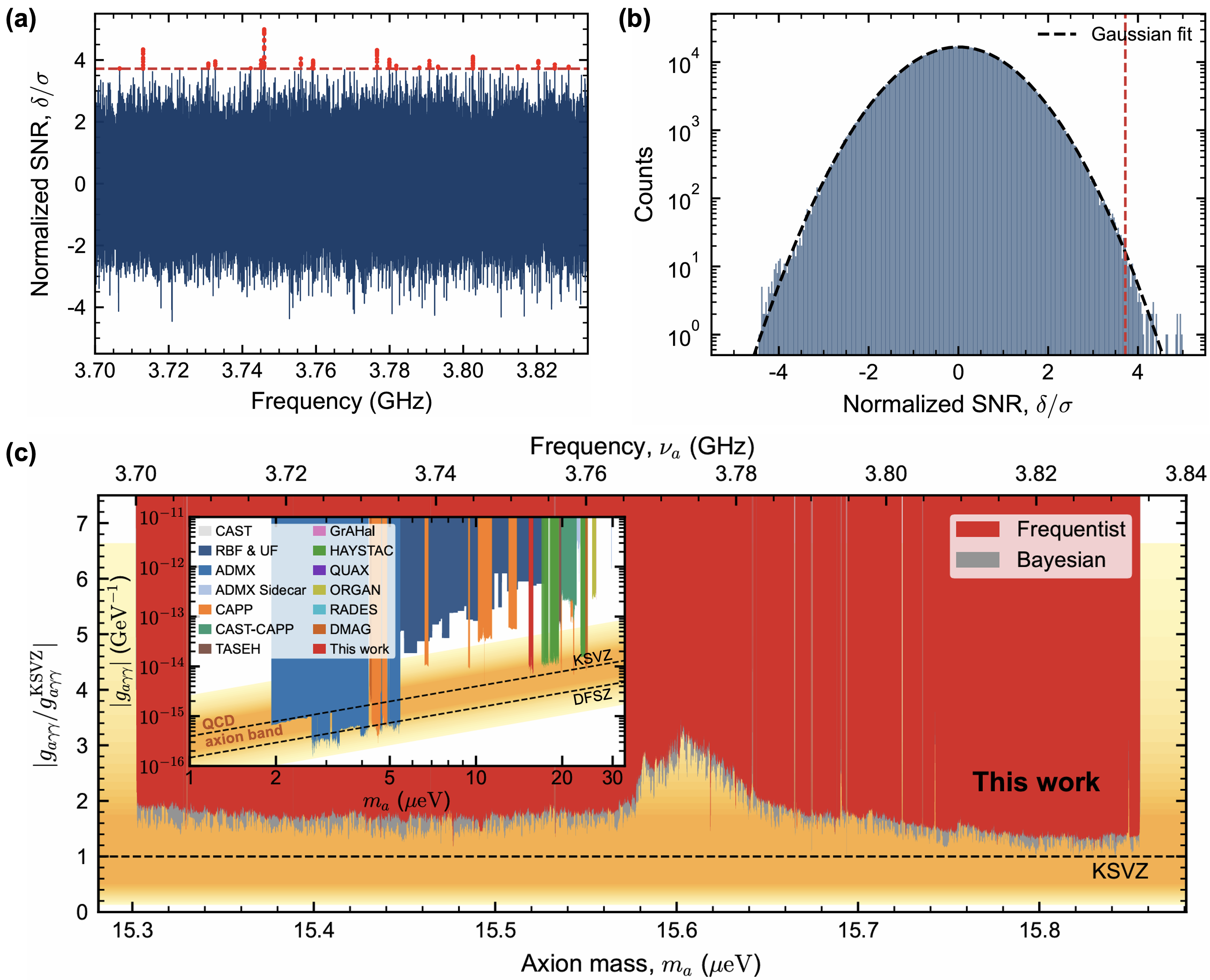}
\caption{Axion dark matter search analysis and physics limits in the mass range of $15.30\text{--}15.85\,\mu\text{eV}$ ($3.700\text{--}3.833\,\text{GHz}$). (a)~Combined grand power spectrum as a function of frequency, showing the $3.718\sigma$ candidate selection threshold (red dashed line) with candidates marked as red dots. (b)~Histogram distribution of normalized SNR~($\delta/\sigma$). The black dashed curve is the fitting to a standard normal distribution, and the red dashed line indicates the selection threshold. (c)~The 90\% confidence-level upper limits on the axion--photon coupling $g_{a\gamma\gamma}$ normalized to the KSVZ coupling. The red and grey shaded regions denote the Frequentist and Bayesian exclusion limits, respectively. The theoretical QCD axion parameter space is highlighted by the gold-shaded band, with the benchmark KSVZ and Dine-Fischler-Srednicki-Zhitnitskii (DFSZ) models represented by black dashed lines. Notably, these limits successfully probe deep into the QCD axion band, reaching the KSVZ benchmark under a Bayesian framework. The inset shows this work along with other axion search results in the extended axion mass range~\cite{CAST2017, RBF1987, UF1990, RBF1989, ADMX2004, *ADMX2018, *ADMX2020, *ADMX2021, *ADMX2025a, *ADMX2025b, ADMXSidecar2018, *ADMXSidecar2023, CAPP2020a, *CAPP2020b, *CAPP2023b, *CAPP2024a, Yoon2022CAPP4, Kwon2021FirstResults, Yi2023DFSZ, Kim2022UltraLowJPA, Bae2024TM020, ahn2024prx, CASTCAPP2022, TASEH2022, GrAHal2021, HAYSTAC2017, *HAYSTAC2018, *HAYSTAC2021, *HAYSTAC2025, QUAX2019, *QUAX2021, *QUAX2023, *QUAX2024, ORGAN2017, *ORGAN2022, *ORGAN2024, *ORGAN2025, RADES2021, *RADES2025, DMAG2025a, *DMAG2026, *DMAG2025b}.}
\label{fig:exclusion}
\end{figure*}

Power spectra were processed through our standard analysis pipeline~\cite{Kwon2021FirstResults, Kim2022UltraLowJPA, Yi2023DFSZ, Bae2024TM020}. Raw spectra were baseline-subtracted using a 4th-order SG filter, yielding a Monte Carlo validated signal efficiency with a band average of $\langle\eta\rangle = 76.04\%$. To eliminate non-physical dips and environmental radio-frequency interference tracking cavity tuning, 19 distinct spurious frequency regions were masked using a $\pm 10\text{-kHz}$ exclusion window. Baseline-subtracted excesses were normalized to the KSVZ axion signal power and vertically combined across overlapping steps. Convolving this combined spectrum with a virialized axion line shape assuming a local dark matter density $\rho_a = 0.45\,\text{GeV/cm}^3$~\cite{Turner1990} yielded the final grand spectrum $\text{SNR}_{\text{grand},j}$ [Fig.~\ref{fig:exclusion}(a)], which precisely follows a standard normal distribution $\mathcal{N}(0,1)$ in the absence of an axion signal [Fig.~\ref{fig:exclusion}(b)].

Applying a selection threshold of $3.718\sigma$ identified 21 candidate clusters. Each candidate was systematically rescanned with extended integration times ($50\text{--}100\,\text{min}$); all were resolved as statistical fluctuations, leaving no persistent axion signature.

We established $90\%$~confidence-level (C.L.) upper limits on the axion--photon coupling $g_{a\gamma\gamma}$. Following the standard haloscope analysis paradigm, our Frequentist framework (target $\text{SNR} = 5.0\sigma$) sets a median exclusion limit of 1.7 times the KSVZ coupling and reaches a peak sensitivity of 1.3 times this benchmark as seen in Fig.~\ref{fig:exclusion}(c). To further contextualize these limits, we additionally evaluated the data using a Bayesian framework ($10\%$~prior update threshold), a statistically rigorous approach previously demonstrated to efficiently extract limits in the presence of haloscope noise fluctuations~\cite{Palken2020BPM}. Under this complementary analysis, the peak sensitivity probes the benchmark KSVZ coupling line ($g_{a\gamma\gamma} \approx g_{\text{KSVZ}}$) [Fig.~\ref{fig:exclusion}(c)]. The localized sensitivity degradation around $15.63\,\mu\text{eV}$ ($3.77\,\text{GHz}$) directly reflects the $T_{\text{sys}}$ spike driven by the JPA mode hybridization [Fig.~\ref{fig:jpa_on}(b)], while the narrow gaps in the exclusion curve correspond to the discrete masked spurious regions.

Systematic uncertainties on $g_{a\gamma\gamma}$ were evaluated following our established framework~\cite{Kwon2021FirstResults, Kim2022UltraLowJPA, Yi2023DFSZ, Bae2024TM020}. Fractional errors propagate in quadrature from the $Y$-factor calibration covariance ($2.2\%$), active noise ratio fluctuations ($0.5\%$), variations in effective quality factor ($5.4\%$), and form-factor tolerance simulations ($\delta C/C = 2.0\%$). Combining these independent sources yields a conservative total systematic uncertainty of $3.1\%$ on $g_{a\gamma\gamma}$.

In summary, we realized a broadly tunable HTS haloscope for an axion dark matter search spanning $15.30\text{--}15.85\,\mu\text{eV}$. By eliminating tuning-induced symmetry-breaking losses via the continuous conductive backing of substrate-stripped REBCO films, the cavity sustained an operational quality factor 3 to 4 times higher than that of copper in an $8\,\text{T}$ magnetic field. This search established robust Frequentist $90\%$~C.L. exclusion limits down to 1.3 times the KSVZ coupling, while a complementary Bayesian analysis achieves KSVZ-level sensitivity. These results demonstrate that high-quality-factor HTS technology effectively compensates for modest magnet volumes ($B^2 V$), offering a practical avenue to accelerate scan rates in high-frequency QCD axion searches.

\textit{Acknowledgments}—This work was supported by the Institute for Basic Science (No. IBS-R040-C1) and by JSPS KAKENHI (Grant No. JP22H04937). A.F.v.L. was supported by a JSPS Postdoctoral Fellowship. 

\textit{Data availability}—The data that support the findings of this study are available from the corresponding author upon reasonable request.

\bibliography{reference}

\end{document}